\RequirePackage[2020-02-02]{latexrelease}
\documentclass{iau}

\usepackage{amsmath}
\usepackage{graphicx}
\usepackage{multirow}

\begin{document}

\lefttitle{Wang et al.}
\righttitle{RRd stars as robust distance and metallicity indicators}

\jnlPage{1}{7}
\jnlDoiYr{2023}
\doival{10.1017/xxxxx}

\aopheadtitle{Proceedings IAU Symposium}
\editors{R. de Grijs, P. A. Whitelock \& M. Catelan, eds.}

\title{Double-mode RR Lyrae star --- robust distance and metallicity indicators}

\author{Shu Wang$^{1,3}$, Xiaodian Chen$^{1,2,3}$, Jianxing Zhang$^{1,2}$, \& Licai Deng$^{1,2,3}$}

\affiliation{$^1$CAS Key Laboratory of Optical Astronomy, National Astronomical Observatories, Chinese Academy of Sciences, Beijing 100101, China \\ email: {\tt shuwang@nao.cas.cn} \\ 
$^2$School of Astronomy and Space Science, University of the Chinese Academy of Sciences, Beijing, 100049, China \\
$^3$Department of Astronomy, China West Normal University, Nanchong, 637009, China
}
   
\begin{abstract}
RR Lyrae (RR Lyr) stars are a well-known and useful distance indicator for old stellar populations such as globular clusters and dwarf galaxies. Fundamental-mode RR Lyr (RRab) stars are commonly used to measure distances, and the accuracy of the determined distance is strongly constrained by metallicity. Here, we investigate the metallicity dependence in the period--luminosity (PL) relation of double-mode RR Lyr (RRd) stars. We find and establish a linear relation between metallicity and period or period ratio for RRd stars. This relation can predict the metallicity as accurately as the low-resolution spectra. Based on this relation, we establish a metallicity-independent PL relation for RRd stars. Combining the distance of the Large Magellanic Cloud and Gaia parallaxes, we calibrate the zero point of the derived PL relation to an error of 0.022 mag. Using RRd stars, we measure the distances of globular clusters and dwarf galaxies with an accuracy of 2-3\% and 1-2\%, respectively. In the future, RRd stars could anchor galaxy distances to an accuracy of 1.0\% and become an independent distance ladder in the Local Group.
\end{abstract}     

\begin{keywords}
RR Lyrae variable stars, RRd variable stars, distance indicators, metallicity, dwarf galaxies
\end{keywords}

\maketitle

\section{Introduction}

RR Lyrae (RR Lyr) stars are low-mass stars that have evolved to the horizontal branch stage and begin to pulsate as they move into the instability strip. The pulsation periods of RR Lyr stars are about 0.2 to 1.0 days. According to the pulsation mode, RR Lyr stars are classified into RRab (fundamental mode), RRc (first-overtone mode) and RRd (double or multiple mode) stars. In the Hertzsprung--Russell diagram, RRd stars are located in the intersection region of the instability strips of RRab and RRc stars.

Most RRd stars are clustered in a sequence on the Petersen diagram \citep{1973A&A....27...89P} and are called classical RRd stars. There are also anomalous and peculiar RRd stars. The anomalous or peculiar RRd stars are located above or below the sequence of classical RRd stars on the Petersen diagram \citep{2021MNRAS.507..781N}. 
The dominant mode of the anomalous or peculiar RRd stars is the fundamental mode, while the dominant mode of classical RRd stars is the first-overtone mode. 
In addition, anomalous or peculiar RRd stars usually have long-term amplitude modulation. In this paper, the default RRd star is the classical RRd star.

RR Lyr stars were searched for by a large number of time domain survey, including the Optical Gravitational Lensing Experiment \citep[OGLE,][]{2009AcA....59....1S, Soszynski2016, Soszynski2019}, the Gaia \citep{Clementini2022}, the Zwicky Transient Facility \citep[ZTF,][]{2020ApJS..249...18C}, the All-Sky Automated Survey for Supernovae \citep[ASASSN,][]{2018MNRAS.477.3145J}, the Asteroid Terrestrial-impact Last Alert System \citep[ATLAS,][]{2018AJ....156..241H}, the Wide-field Infrared Survey Explorer \citep[WISE,][]{2018ApJS..237...28C}, and the PanSTARRS1\citep{2017AJ....153..204S}.

RR Lyr stars are good distance tracers for the old population. They satisfy the metallicity--luminosity (LZ) relation in the $V$ band and the period--metallicity--luminosity (PLZ) relation in infrared bands. 
Since the two periods of RRd stars also satisfy the PLZ relation of RRab or RRc, respectively, RRd stars also have LZ and PLZ relations. Usually, RRd stars were not adopted to measure distances because the determined periods of RRd stars are not as accurate as those of RRab stars.

For RRd stars, both theory \citep{Popielski2000, Marconi2015} and observations \citep{Catelan2015, Soszynski2019} suggest that compared to metal-poor RRd stars, metal-rich RRd stars have shorter fundamental periods and smaller period ratios. 
There seems to be a good linear relation between the period ratio and metallicity \citep{Braga2022}. 
In this work, we investigate the metallicity dependence in the period--luminosity (PL) relation of RRd stars and establish their PL relation \citep{Chen2023}.

\section{Period--Metallicity Relation}

We collect RRd stars mainly based on Gaia and OGLE databases. From Gaia DR3, we obtain 1021 Galactic RRd stars. From OGLE, we collect 2083 and 674 RRd stars belonging to the Large Magellanic Cloud (LMC) and the Small Magellanic Cloud (SMC), respectively. 
Metallicity information for these stars is taken from low-resolution spectra of the Sloan Digital Sky Survey \citep[SDSS,][]{Eisenstein2011} and the Large Sky Area Multi-Object Fiber Spectroscopic Telescope \citep[LAMOST,][]{Cui2012}. 
After cross-matching, we obtain 100 RRd stars with SDSS or LAMOST metallicities. 
In addition, we supplement another 203 RRd stars with SDSS or LAMOST metallicity based on the ZTF data.
The final sample contains 207 and 96 RRd stars with metallicity measurements from SDSS and LAMOST, respectively.

\begin{equation}\label{equ1}
 \begin{aligned}
   {\rm [Fe/H]_{SDSS}} = & -(173\pm38)(\log \frac{P_{\rm 1O}}{P_{\rm F}}-\log0.745) \\
   &-(6.62\pm0.88)(\log P_{\rm 1O}-\log0.37) -(1.76\pm0.01),  \\
   {\rm [Fe/H]_{LAMOST}} = & -(184\pm70)(\log \frac{P_{\rm 1O}}{P_{\rm F}}-\log0.745) \\
   &-(5.69\pm1.50)(\log P_{\rm 1O}-\log0.37)-(1.68\pm0.03),  \\
 \end{aligned}
\end{equation}

Based on the selected RRd stars, we establish the period--period ratio--metallicity relation. 
The results are shown in equation 1.  
For the SDSS metallicity, the scatter $\sigma$ is 0.16 dex and the correlation coefficient $R^2$ is 0.74. 
For the LAMOST metallicity, the scatter $\sigma$ is 0.21 dex and the correlation coefficient $R^2$ is 0.60. 
The scatter of this relation decreases when a tighter criterion on metallicity error is adopted. 
For example, when we restrict the metallicity error to less than 0.04, with the rest of 56 RRd, the scatter of this relation will decrease to 0.13 dex.
By comparing the scatter of the period--period ratio--metallicity relation with the external error of metallicity, we find that the metallicity estimated from the period and the period ratio of RRd stars can be as accurate as the low-resolution spectra. 

With metallicities from high-resolution spectra, the scatter of the period--period ratio--metallicity relation could be greatly reduced. 
For example, for Cepheids, the scatter of the relation is only 0.033 dex \citep{Kovtyukh2016}.
In the future, with metallicities from high-resolution spectra, the period--period ratio--metallicity relation of RRd stars could also have a small scatter. \cite{Kovtyukh2016} found that for double--mode Cepheids, their metallicity depend on period ratio more than the period. 
Based on our sample, we find that RRd metallicity is more dependent on the period than the period ratio.

\begin{figure*}[ht]
\centering
\vspace{-0.0in}
\includegraphics[angle=0,width=4.0in]{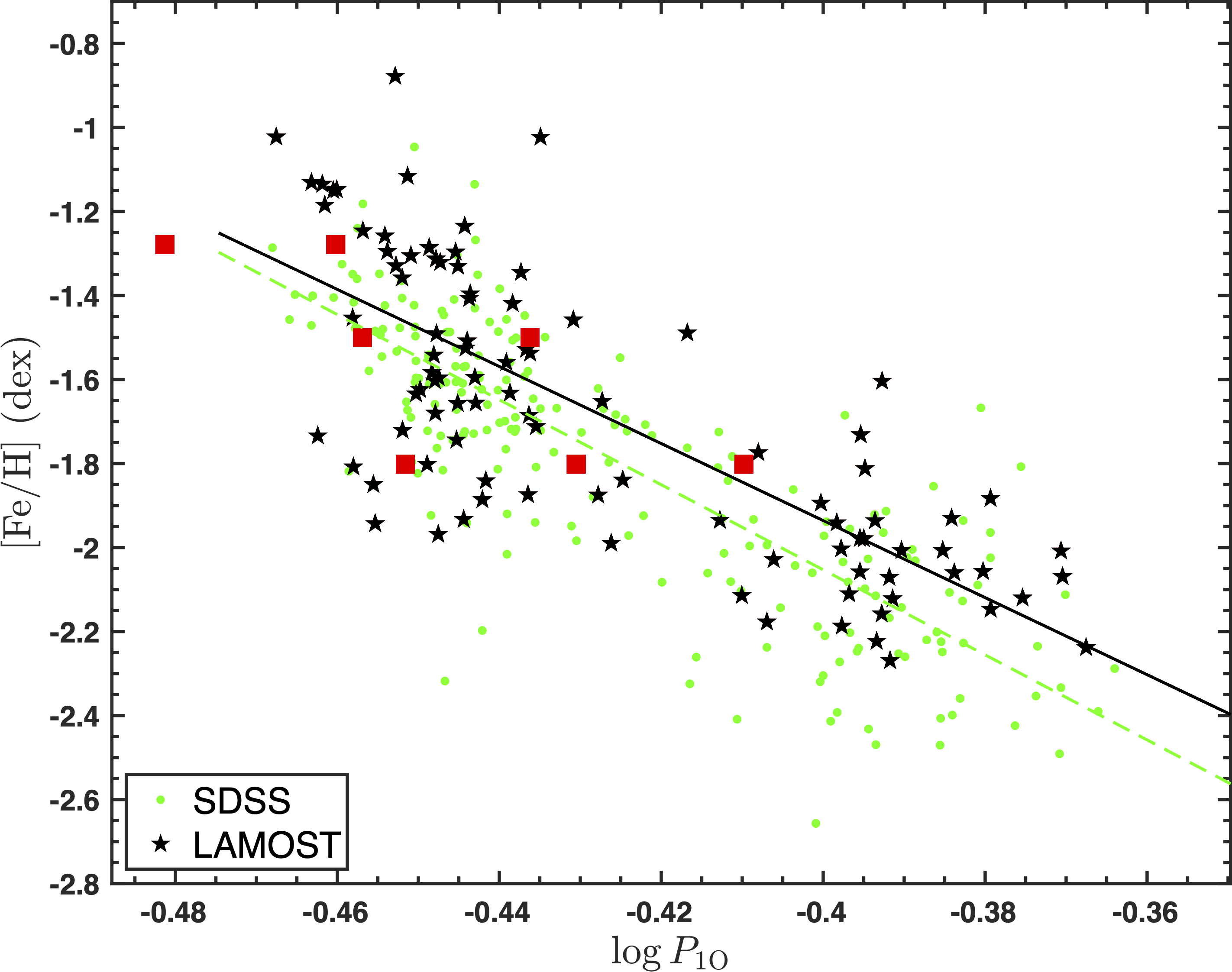}
\vspace{-0.0in}
\caption{The Period--metallicity diagram of RRd stars with metallicities from SDSS (green dots) and LAMOST (black pentagrams). The green dashed line and the black solid line are linear fit lines. Red squares are the grid points of the theoretical model.}
\label{fig:PZ} 
\end{figure*}

Figure 1 shows the distribution of RRd stars in the period--metallicity (PZ) diagram. The green dots and black pentagrams are RRd stars with metallicities from SDSS and LAMOST, respectively. Red squares are the grid points of the theoretical model \citep{Marconi2015}. 
There is an intuitive linear relationship between the metallicity and the period. The green dashed line and the black solid line are linear fit lines. 
The determined PZ relations for RRd stars with SDSS metallicities are listed in equation 2.

\begin{equation}\label{equ2}
\begin{split}
  {\rm [Fe/H]_{SDSS}} = (-10.33\pm0.47) \log P_{\rm F}+(-4.87 \pm 0.14),  \sigma=0.17 \, {\rm dex}, R^2=0.71, \\
  {\rm [Fe/H]_{SDSS}} = (-10.15\pm0.46) \log P_{\rm 1O}+(-6.11 \pm 0.19),  \sigma=0.17 \, {\rm dex}, R^2=0.71
\end{split}
\end{equation}

\section{Period--Luminosity Relation} 

Combining the PLZ relation with the period--period ratio--metallicity relation, the period--period ratio--luminosity relation is derived. We determined the period--period ratio--luminosity relation based on OGLE LMC RRd stars. We find that there is an approximate linear relation between the period and the period ratio, with a correlation coefficient of 0.7 and a dispersion of 0.0003.
Therefore, we try to simplify the period--period ratio--luminosity relation to the period--luminosity (PL) relation. 
We find that the absolute magnitude difference predicted by the PL relation and the period--period ratio--luminosity relation is small.
The scatter in magnitude difference is 0.01 mag and the maximum magnitude difference is only 0.06 mag.
In addition, the scatter of the PL relation and the period--period ratio--luminosity relation is the same. 
Since the metallicity is more dependent on period, we ultimately simplify the period--period ratio--luminosity relation to the PL relation (see equation 3). Here, we adopt the Wesenheit magnitude \citep{1982ApJ...253..575M} i.e. $W_{VI}=I-1.55(V-I)$ and $W_{G, BP, RP}= G -1.90(BP - RP)$.

\begin{equation}\label{equ3}
 \begin{aligned}
  &M_{W_{VI}} = (-4.523\pm0.156) \log P_{\rm F}+(16.620 \pm 0.048) - {\rm DM_{LMC}}, \sigma=0.132 \,{\rm mag}, \\
  &M_{W_{VI}} = (-4.434\pm0.153) \log P_{\rm 1O}+(16.079 \pm 0.067) - {\rm DM_{LMC}}, \sigma=0.132 \,{\rm mag}, \\
  &M_{W_{G, BP, RP}} = (-3.623\pm0.229) \log P_{\rm F}+(17.042 \pm 0.071) - {\rm DM_{LMC}}, \sigma=0.159 \,{\rm mag}, \\
  &M_{W_{G, BP, RP}} = (-3.557\pm0.225) \log P_{\rm 1O}+(16.606 \pm 0.098) - {\rm DM_{LMC}}, \sigma=0.159 \,{\rm mag}
 \end{aligned} 
\end{equation}

We combine the LMC distance \citep{2019Natur.567..200P} and Gaia parallaxes to calibrate the zero points of the PL relations as $M_{W_{G, BP, RP}}(P_{\rm 1O}=0.37 {\rm d})=-0.348\pm0.022$ mag and $M_{W_{VI}}(P_{\rm 1O}=0.37 {\rm d})=-0.496\pm0.022$ mag. Meanwhile, we derive the zero point of the Gaia DR3 parallax, which is $13.4\pm5.8\ \mu{\rm as}$. This parallax zero point agrees well with that determined by classical Cepheids \citep{2021ApJ...908L...6R}.

\section{Distance and Metallicity Measurements}

\begin{figure*}[ht]
\centering
\vspace{-0.0in}
\includegraphics[angle=0,width=4.0in]{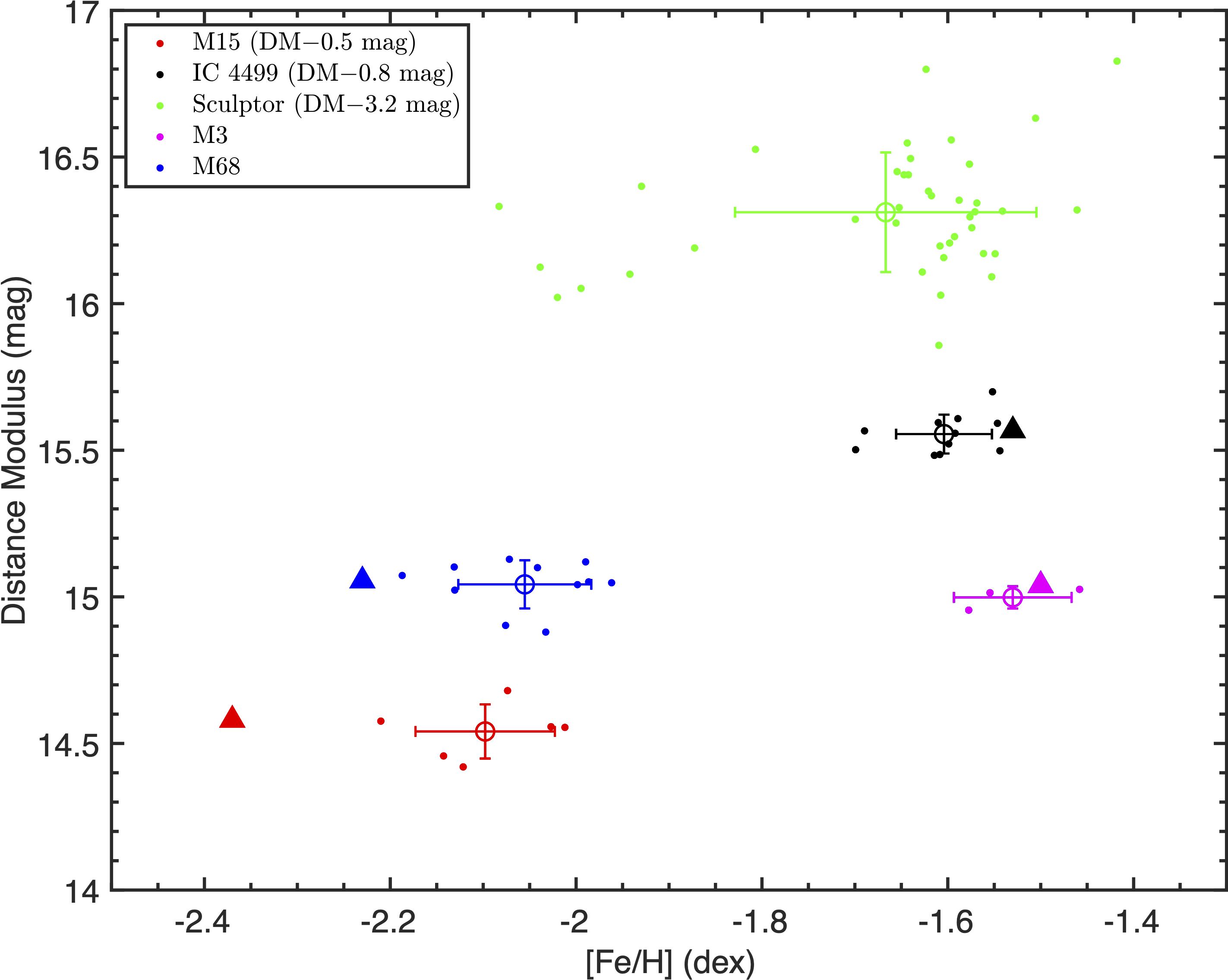}
\vspace{-0.0in}
\caption{Metallicity and distance modulus (DM) determinations based on RRd stars for M15 (red dots), IC 4499 (black dots), Sculptor (green dots), M3 (magenta dots), and M68 (blue dots). The mean values of RRd stars in five targets are shown as open circles with $1\sigma$ internal errors. Solid triangles indicate the parameters of the four globular clusters from the Harris's globular cluster catalog. To make the figure clearer, the distance modulus of M15, IC 4499, and Sculptor are reduced by 0.5 mag, 0.8 mag, and 3.2 mag, respectively.}
\label{fig:distance} 
\end{figure*}

The established PL relation and PZ relation of RRd stars can be used to determine the distances and metallicities of distant objects.
We measure the distances and metallicities of four globular clusters (IC 4499, M15, M3, and M68) and a dwarf galaxy (Sculptor) based on RRd stars with Gaia data. 
Figure 2 shows the derived metallicities and distance moduli. The different colored dots indicate individual metallicity and distance modulus determinations based on RRd stars in M15 (red), IC 4499 (black), Sculptor (green), M3 (magenta), and M68 (blue). 
Solid triangles indicate the results for the four globular clusters from the globular cluster catalog of \cite{Harris1996}.  
The mean values of RRd stars in five targets are shown as open circles with $1\sigma$ internal errors. 
For dwarf galaxy Sculptor, the distance accuracy is about 1.8\%, and the metallicity dispersion is 0.2 dex. 
For globular clusters, the distance accuracy is about 2-3\%, and the metallicity dispersion is 0.05 dex.
Recently, we use RRd stars to measure the distances and metallicities of other dwarf galaxies. 
For example, we measure the distances and metallicities of five dwarf galaxies in M31, where the distance accuracy is about 1.8\%.
We also determine the distance of the Fornax dwarf galaxy with an accuracy of 1.2\% based on $\sim130$ RRd stars.

There are several advantages of adopting RRd stars as distance indicators. Since the RRd star PL relation is not affected by metallicity, the galaxy distance error can be optimized to 1.2\% based on $\sim100$ RRd stars. This accuracy is as good as the distance accuracy of the LMC. RRd stars are not rare, and they account for 3-5\% of RR Lyr stars. The number of RRd stars in the SMC, LMC, and the Galactic halo is comparable to the total number of classical Cepheids in these galaxies. Moreover,  distance measurements based on RRd stars are less affected by extinction and binarity than the classical Cepheids. RRd stars are the best distance indicators for dwarf galaxies. They can also provide metallicity information for distance measurements of RRab stars. 
In the future, the China Space Station Telescope and the Vera C. Rubin Observatory Legacy Survey of Space and Time will discover more RRd stars. The RRd star can be an independent distance ladder in the Local Group, used to check the Cepheid \citep{2022ApJ...934L...7R} and the tip of the red giant branch \citep{2019ApJ...882...34F} distance ladders.

\section{Summary}

We discover a period--(period ratio)--metallicity relation for RRd stars and find it can predict metallicity as accurately as the low-resolution spectra. 
We propose that the PL relation of RRd stars is not affected by the metallicity and derive the PL relation of RRd stars. 
The zero point of the PL relation is determined with an error of 0.022 mag (1.0\% error in the distance). 
We find that RRd stars can anchor galaxy distances with an accuracy of 1-2\% (dwarf galaxies) and 2-3\% (globular clusters). 
In the Future, RRd stars will be established as an independent distance ladder in the Local Group. 

\section*{Acknowledgements}
This work is supported by the National Natural Science Foundation of China (NSFC) through the projects 12003046, 12173047, 11903045, 12133002, and 11973001. 
S.W. and X.C. acknowledge support from the Youth Innovation Promotion Association of the CAS (grant No. 2023065 and 2022055).

\end{document}